\documentclass{desyproc}

\usepackage{graphicx}
\usepackage{amssymb}

\begin{document}
\title{
The Sun in Hidden Photons}

\author{{\slshape Javier~Redondo}\\[1ex]
Max Planck-Institute f\"ur Physik, F\"ohringer Ring 6, D-80805 M\"unchen, Germany}

\contribID{redondo\_javier}

\desyproc{DESY-PROC-2011-04}
\acronym{Patras 2011} 
\doi  


\maketitle

\begin{abstract}
We present some aspects and first results of the emission of sub-eV mass hidden photons from the Sun. The contribution from a resonant region below the photosphere can be quite significant, raising previous estimates. This is relevant for the Telescope for Hidden Photon Search, TSHIPS I, currently targeting at meV-mass hidden photons with ${\cal O}(10^{-6})$ kinetic mixing with the photon. These particles could account for the large  effective number of  neutrinos pointed at by the cosmic microwave background and other large-scale structure probes, and are motivated in some scenarios of string theory. 
\end{abstract}


Paraphotons were first discussed by L.~Okun as means to study the limits of our knowledge on electrodynamics~\cite{Okun:1982xi}. 
The basic idea is that the photon (the interaction state that couples to electric charge) might not be a propagation eigenstate if there is a new species of vector boson that mixes with it. Okun parametrized the precision achieved in many observables in electrodynamics by means of the mixing angle, $\chi$, and the mass, $m_{\gamma'}$ of the non-zero mass eigenstate, which he called paraphoton. 
Interestingly, he found that the most astonishing constraints on the mixing angle do not come from virtual effects as Lamb-shifts or deviations of Coulomb's law, but from the real emission of paraphotons from the Sun, although this depends strongly on  $m_{\gamma'}$. 
In following works, the solar constrains were refined and also the idea of detecting the flux of paraphotons from the Sun with an helioscope was considered experimentally~\cite{Popov:1991}. 
In these studies it was already realized that aiming at the low-energy part of the HP solar spectrum was more advantageous than focusing in the X-ray regime. 

Some years later, theoretical advances put these ideas in a different context. 
It was realized that similar particles arise in extensions of the standard model as gauge bosons of hidden U(1) symmetries, i.e. symmetries under which fields of the standard model are left unchanged. The paraphoton was re-baptized as ``hidden photon" (HP) in this context.  Despite the hidden nature of these particles at tree-level, the key ingredient, the mixing with the photon can happen at one-loop level via radiative corrections. These corrections mix the photon and HP field strengths, $F_{\mu\nu}$ and $F'_{\mu\nu}$ resp., contributing to the kinetic mixing operator~\cite{Holdom:1985ag}, 
\begin{equation}
\mathcal{L}_{mix}=-\frac{1}{4}\chi F_{\mu\nu}F^{\prime\mu\nu}\; ,
\end{equation}
which being a renormalizable operator, it is only logarithmically sensitive to the energy-scale at which mixing is generated. This means that irrespectively of the high-energy physics completion of the SM that connects with the hidden sector, one can expect mixings of the order of a radiative correction $\chi\sim 10^{-3}$~\cite{Holdom:1985ag}. 
These relatively large couplings are already excluded by a number of experiments and observations in the HP mass range $10^{-14}\sim 10^{10}$ eV~\cite{Jaeckel:2010ni}. But models in which the hidden and/or the electromagnetic U(1) are embedded in a non-abelian group, on in which the hidden gauge coupling is small---such as in the LARGE volume scenarios of Type-IIb string theory--- predict necessarily smaller mixings~\cite{Goodsell:2009xc}. Values down to $\chi\sim 10^{-12}$ arise naturally in string theory, see~\cite{Goodsell:2010ie} for a review.

More importantly, hidden photons have been motivated by recent observations. 
In cosmology they can account for the cold dark matter of the universe either via the misalignment mechanism ---in much the same fashion as axions (see~\cite{Nelson:2011sf,Arias:2012mb} and the contribution of P.~Arias to these proceedings)--- or through their resonant production in the early universe~\cite{Redondo:2008ec} (for $m_{\rm \gamma'}\sim 100$ keV, $\chi \sim 10^{-12}$). 
The resonant production of $m_{\rm \gamma'}\sim $ meV HPs with $\chi\sim 10^{-6}$ happens after BBN and before structure formation. The creation of HPs and depletion of photons implies a sizable number of extra effective neutrinos $\Delta N_{\rm eff}$ (a measure of the amount of dark radiation) imprinted in the CMB~\cite{Jaeckel:2008fi}, just as recent cosmological probes are suggesting~\cite{Komatsu:2010fb}. 
This last region of parameter space is the primary goal of the first Telescope for Solar Hidden Photon Search, TSHIPS I (see the contribution of M.~Schwartz to these proceedings). In order to interpret its results it is crucial to have a reliable estimate of the solar HP flux in the visible region, where the telescope sensitivity is optimized.   

The solar HP flux comes from oscillations of the photons of the solar interior into HPs, which then escape unimpeded, see~\cite{Redondo:2008aa} for a detailed theoretical study. In an inhomogenous medium, the oscillation probability can be estimated in a perturbative fashion through an integral over the putative photon trajectory inside the Sun, $r=r(s)$, as 
\begin{equation}
P(\gamma\to\gamma')=\left|\chi \frac{m_{\rm \gamma'}^2}{2\omega}\int_0^\infty  e^{i \Phi(s)-\tau(s)}ds\right|^2
\end{equation}
where $\omega$ is the photon energy and 
\begin{equation}
\Phi(s)=\int_0^s \frac{m_\gamma^2(r(s'))-m_{\rm \gamma'}^2}{2 \omega}ds' \quad ; \quad  \tau(s)=\int_0^s \frac{\Gamma(r(s'))}{2}ds' .
\end{equation}
Here $M^2=m_{\rm \gamma}^2-i \omega\Gamma$ is the effective photon mass in the medium, generally complex and depending on the solar radial coordinate $r$. 
If it does not change appreciably in an absorption length, $\Gamma |d \log |M^2|/ d r|^{-1} \ll 1$, we can approximate $M^2(r)\simeq M^2(r_0)$ to obtain  
\begin{equation}
\label{prob}
P(\gamma\to\gamma')  = \frac{\chi^2 m_{\rm \gamma'}^4}{(m^2(r_0)-m_{\rm \gamma'}^2)^2+ (\omega \Gamma(r_0))^2} . 
\end{equation}
The differential HP flux results from integrating over the solar interior the photon emission rate times the conversion probability, which can be written and parametrized as 
\begin{equation}
\nonumber
\frac{d \Phi_{\gamma'}}{d\omega} = \int_0^{R_\odot} \frac{r^2 d r}{D^2_{\rm Sun}}  \frac{\Gamma}{e^{\omega/T}-1}\frac{\omega^2}{\pi^2}  P(\gamma\to \gamma') 
\equiv \chi^2\left(\frac{m_{\gamma'}}{\rm eV}\right)^4 \int_0^{R_\odot}dr F(r,\omega,m_{\gamma'})
\end{equation}
where $T=T(r)$ is the temperature and $D_{\rm Sun}\simeq 1.5\times 10^{8}$ km. 
A model of photon refraction and absorption of the solar interior $M^2(r)$ is needed. 
Refraction is determined by forward scattering off the ambient electrons, either free or bound in atoms, while absorption happens through a number of reactions involving electrons, mainly inverse bremsstrahlung and Thomson scattering.  
The required quantities depend on the position inside of the Sun via $T$, the density of electrons ($n_e$) and the solar composition, which are available from solar models. The calculations shown here use data from~\cite{TurckChieze:2001ye}, which offers great detail in the outer layers of the Sun. 

\begin{figure}[t]
   \centering
   \includegraphics[width=4.5cm]{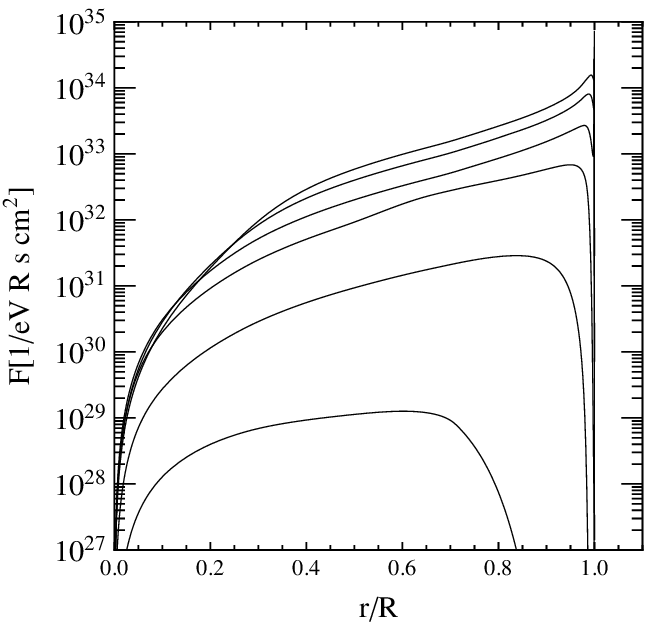} 
       \includegraphics[width=4.5cm]{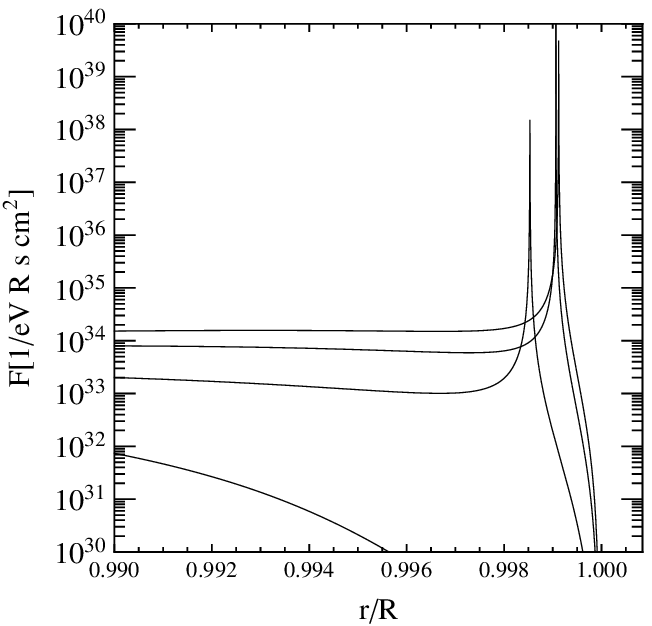} 
       \includegraphics[width=4.5cm]{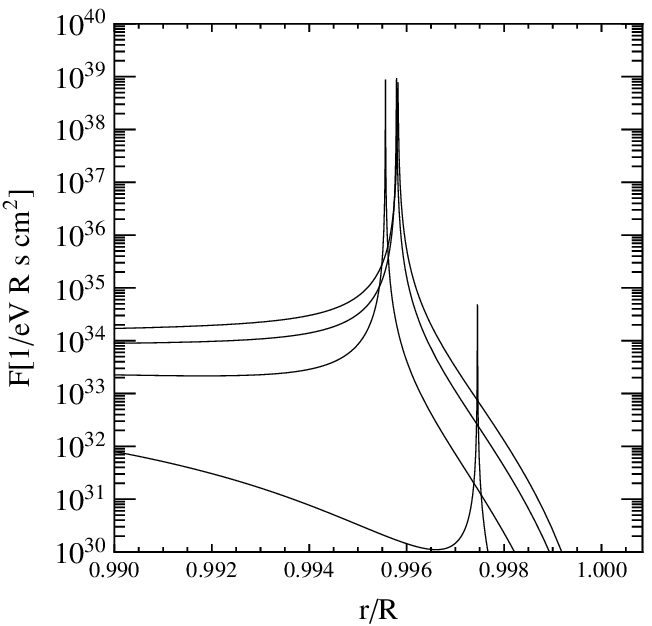} 
   \caption{
   \small The flux function $F(r,\omega,m_{\gamma'})$ for $m_{\gamma'}= 0$ (LEFT and CENTER) and  $m_{\gamma'}= 0.1$ eV (RIGHT). The curves show different HP energies $\omega=2,3,6,14,10^2$ and $10^3$ eV (top to bottom). In the last plots the two highest energies are not visible. }
   \label{fig:F}
\end{figure}

Before discussing our numerical results, it is possible to obtain a bit of analytical insight. For the low mass HPs of interest, the flux function $F$ does not depend on $m_{\gamma'}$ in the solar interior $F(r,\omega,m_{\gamma'})\simeq F(r,\omega,0)$. In this region, $m_\gamma^2$ is determined by the electron density and mass ($m_e$) through the plasma frequency, $m_\gamma^2=\omega_{\rm P}^2 \propto n_e/m_e$, while the absorption coefficient is dominated by electron-proton bremsstrahlung, which gives $\Gamma\propto n_e n_p (1-e^{-\omega/T})/(m_e^{3/2} T^{1/2} \omega^3)$ where $n_p$ is the proton density. Using Eq.~\ref{prob} with $n_p\simeq n_e$ and $\omega\Gamma\ll m_\gamma^2$ we find 
\begin{equation}
F(r,\omega)\propto r^2 e^{-\omega/T}/(\omega\sqrt{T})  . 
\end{equation} 
This shows that the flux of low mass HPs from the solar interior is not suppressed by the larger electron density, as previously stated. 
It is however suppressed by the temperature
\begin{wrapfigure}{r}{0.5\textwidth}
   \centering
\vspace{-.2cm} 
   \includegraphics[width=6.5cm]{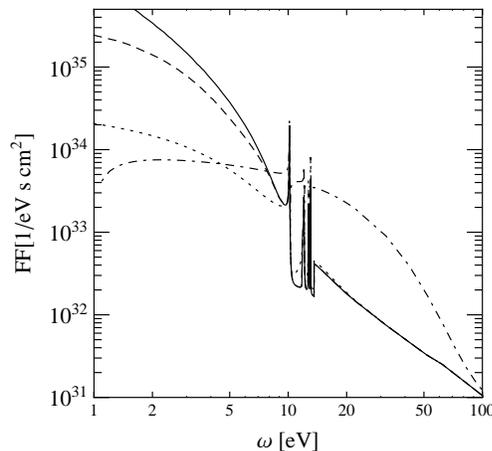} 
\vspace{-.3cm}  
 \caption{
   \small The Sun-integrated HP flux function $\int_0^{R_\odot} dr F(\omega,m_{\gamma'})$ as a function of the HP energy, for $m_{\gamma'}= 0,0.01,0.1$ and $1$ eV, resp. solid, dashed, dotted and dot-dashed. }
   \label{fig:FF}
\vspace{-.3cm}     
\end{wrapfigure}
 and the volume factor $r^2$ such that the solar external shells and low HP energies do indeed dominate. 
The energy emission is more evenly distributed, as the integral over energies of $F$ is $\propto r^2\sqrt{T}$, which is relatively constant inside the Sun. 
A plot of $F(r,\omega,0)$ is shown in Fig.~\ref{fig:F} (LEFT). 
In this figure we also see sharp peaks close to the solar surface, which we have magnified in Fig.~\ref{fig:F} (CENTER).  These peaks arise from resonances where $m_\gamma^2=0$ because the positive contribution from free electrons is cancelled by a negative contribution arising from electrons bounded in Hydrogen atoms oscillating below their resonant energy. In the Sun, the most important resonance is the Ly-$\alpha$ transition at $\omega_{12}\simeq 10.2$ eV. Other resonances play a role only for HP energies very close to them. 

For nonzero HP mass,  the resonant condition becomes $m_\gamma^2-m_{\gamma'}^2=0$ and the regions of resonant conversion move deeper into the Sun as $m_\gamma^2$ grows more positive inside the Sun where the density and the ionization fraction are larger, see~\cite{Cadamuro:2010ai}. 
In Fig.~\ref{fig:F} (RIGHT) we show how the peak structure of $F$ moves inwards in the $m_{\gamma'}=0.1$ eV case with respect to the massless limit.

At low energies, the resonant photon$\leftrightarrow$HP conversion contributes significantly to the total HP flux or dominates it completely. In Fig.~\ref{fig:FF} we show the Sun-integrated emission for different HP masses as a function of energy. The results are significantly larger than the conservative estimate used in~\cite{Gninenko:2008pz}, based upon an uncomplete solar model and neglecting bound electrons.   
In this calculation we have only taken into account the resonant transitions of Hydrogen and assumed the validity of Eq.~\ref{prob}. The contribution of Helium and metals is likely to be irrelevant for the low energy HP flux ($\omega\sim {\cal O}$(eV)) since they are much less abundant than Hydrogen. Also, their strongest resonant transitions happen at frequencies higher than $\omega_{12}$ and their contributions to  $m_\gamma^2$ are inversely proportional to the resonant frequency squared. The assumption of Eq.~\ref{prob} breaks necessarily down when approaching the solar surface. However, most of the low energy emission comes from the resonance which happens at a finite depth even in the $m_{\gamma'}=0$ case so Eq.~\ref{prob} provides a good estimate in most of the cases. 
We plan to refine these estimates and arguments in a further publication. 

\section*{Acknowledgments}
I am very thankful to Davide Cadamuro, J\"oerg J\"aeckel, Georg Raffelt, Andreas Ringwald, Pat Scott and Sergey Troitsky for discussions around the topic presented here and to the organizers of the 7th Patras meeting  for a wondrous time in a much inspiring atmosphere. 


\begin{footnotesize}

\end{footnotesize}


\end{document}